\documentclass[submission,copyright,creativecommons]{eptcs}
\providecommand{\event}{LTT 2026} % Name of the event you are submitting to

\usepackage{iftex}

\ifpdf
  \usepackage{underscore}         % Only needed if you use pdflatex.
  \usepackage[T1]{fontenc}        % Recommended with pdflatex
\else
  \usepackage{breakurl}           % Not needed if you use pdflatex only.
\fi

\usepackage{amssymb,amstext,amsmath,amsthm}
\newtheorem*{proposition*}{Proposition}%[theorem]

\newcommand{\ERASER}[1]{}

\newcommand{\Id}{\mathsf{Id}}

\newcommand{\id}{\mathsf{id}}
\newcommand{\NN}{\mathsf{N}}

\newcommand{\UU}{\mathsf{U}}

\newcommand{\level}{\mathsf{level}}

\newcommand{\app}[2]{{#1\,#2}} % many applications still hard-coded with ~

\newcommand{\Ctx}{\mathrm{Ctx}}
\newcommand{\Sub}{\mathrm{Hom}}
\newcommand{\Ty}{\mathrm{Ty}}
\newcommand{\Tm}{\mathrm{Tm}}
\newcommand{\Lctx}{\mathrm{Lctx}}
\newcommand{\Lsub}{\mathrm{Lhom}}
\newcommand{\Ltm}{\mathrm{Ltm}}
\newcommand{\op}{\mathrm{op}}

\def\lhom{\mathrm{lhom}}

\newcommand{\refl}{\mathsf{r}}

\def\NN{\mathsf{N}}
\def\Nbb{\mathbb{N}}
\def\UU{\mathsf{U}}

\newcommand{\N}{\mathsf{N}}

\newcommand{\T}{\mathsf{T}}

\def\Pihat{\Pi}

\def\Obj{\mathrm{obj}}
\def\sub{\mathrm{hom}}
\def\id{\mathrm{id}}
\def\lHom{\mathrm{lhom}}
\def\lctx{\mathrm{lctx}}

\def\ltm{\mathrm{ltm}}

\def\leq{\mathrm{leq}}

\def\lp{\mathrm{lp}}
\def\lq{\mathrm{lq}}
\def\s{\mathrm{s}}
\def\lid{\mathrm{lid}}

\newcommand{\ctx}{\mathrm{ctx}}
\newcommand{\ty}{\mathrm{ty}}
\newcommand{\tm}{\mathrm{tm}}
\newcommand{\tuple}[1]{\langle #1 \rangle}

\newcommand{\cext}{.}
\def\p{\mathrm{p}}
\def\q{\mathrm{q}}
\def\app{\mathsf{app}}
\def\U{\mathsf{U}}
\def\T{\mathcal{T}}
\newcommand{\Ta}{\mathrm{T}}

\def\L{{\mathcal{L}}}

\def\CwF{\mathrm{CwF}}

\def\CwFext{\mathrm{CwF_{tower}}}
\def\CwFint{\mathrm{CwF_{up}}}
\def\LCwF{\mathrm{LCwF}}
\def\LCwFint{\LCwF_{\mathrm{up}}}
\def\Fam{\mathrm{Fam}}

\def\C{\mathcal{C}}
\def\W{\mathsf{W}}

\def\Sigmaext{{\Sigma^\mathrm{tower}}}
\def\Sigmaint{{\Sigma^\mathrm{up}}}
\def\TText{{\mathbf{TT}^\mathrm{tower}}}
\def\TTint{{\mathbf{TT}^\mathrm{up}}}
\def\Text{\T_\mathrm{tower}}
\def\Tint{\T_\mathrm{up}}

\title{A Generalized Algebraic Theory\\ for Type Theory
with Explicit Universe Polymorphism}

\author{Marc Bezem\institute{University of Bergen}\email{Marc.Bezem@uib.no}\\
\and Thierry Coquand \quad\quad\quad\quad Peter Dybjer\institute{Chalmers University of Technology and University of Gothenburg}
\email{Thierry.Coquand@cse.gu.se \quad\quad peterd@chalmers.se}
\and Mart\'{\i}n Escard\'o\institute{University of Birmingham}
\email{m.escardo@bham.ac.uk}
}

\def\titlerunning{A Generalized Algebraic Theory for Type Theory with Explicit Universe Polymorphism}
\def\authorrunning{Marc Bezem, Thierry Coquand, Peter Dybjer \& Mart\'{\i}n Escard\'o}
\def\copyrightholders{M. Bezem, Th. Coquand, P. Dybjer \& M. Escard\'o}

\begin{document}
\maketitle

\begin{abstract}
We present generalized algebraic theories corresponding to slightly modified versions of two of the type theories in our paper
{\em Type Theory with Explicit Universe Polymorphism}. We first present a generalized algebraic theory for categories with families with extra structure corresponding to Martin-Löf type theory with an external tower of universes. We then present a generalized algebraic theory for level-indexed categories with families with extra structure corresponding to Martin-Löf type theory with explicit universe polymorphism: a theory with universe level judgments, internally indexed universes, and level-indexed products. In this way we get abstract characterizations of the two theories as initial models of their respective generalized algebraic theories. We thus abstract from details of the grammar and inference rules of the type theories and highlight their high-level structure. More broadly, the present work can be viewed as a case study of a uniform approach to categorical logic based on generalized algebraic theories and categories with families. We also discuss the relevance to Voevodsky's initiality conjecture project. \ERASER{\todo[inline]{PD: Add something about level equality sorts? Maybe this is too specific to be mentioned in the abstract? It could be "an interesting feature of the gat for explicit universe polymorphism is that we employ level equality sorts for encoding an equational constraint in the type of codes for universes"? }}
\end{abstract}

\section{Introduction}

In our paper \cite{BezemCDE22} on type theory with explicit universe polymorphism, we proposed several extensions of Martin-Löf type theory with universe polymorphism. We followed Courant's approach \cite{Courant02} and added universe level {\em judgments}:
$$
l\ \level
\hspace{5em}
l = l'
$$
to the usual judgment forms of type theory. Moreover, all judgments may depend on universe level variables as well as ordinary variables declared in the context. We emphasized that universe levels do {\em not} form a {\em type} in our setting, and instead we added the above judgment forms\footnote{In Agda universe levels form a type, although recently Agda has introduced, in version 2.6.4, an option for disabling universe levels forming a type~\cite{agda:leveluniv}). However, it is unclear whether this option results in the same theory as ours.}.  To regain some of the lost expressivity (in a controlled way), we added level-indexed products $[\alpha]A$ of families of types $A\ (\alpha\ \level)$ to our theory.

In this paper we provide alternative presentations of (slightly modified versions of) two of the theories in the above-mentioned paper: Martin-Löf type theory with an external tower of universes $\TText$ and Martin-Löf type theory with explicit universe polymorphism $\TTint$ as outlined in the previous paragraph. We present the corresponding {\em generalized algebraic theories (gats)} $\Sigmaext$ and $\Sigmaint$. Gats were introduce by Cartmell \cite{cartmell:phd,cartmell:apal} as a generalization of many sorted algebraic theories where sort symbols and operator symbols may have dependent types.
%Our aim is to characterize
%$\TText$ and $\TTint$ as initial models of $\Sigmaext$ and $\Sigmaint$ respectively, provided we decorate some terms (such as applications) with types.

In Section \ref{sigmaext} we will present $\Sigmaext$, the infinitary gat of categories with families (cwfs)  \cite{dybjer:torino} with extra structure for the small type formers ($\Pi, \Sigma, \N_0, \N_1, \N, \W, \Id$) of Martin-Löf type theory and for the tower of universes $\U_l$ indexed by external natural numbers $l \in \NN$. Cwfs are models of the basic rules of dependent type theory: context formation, context morphism formation, substitution in types and terms, projection morphism, assumption.  (Note that we use $\Sigma$ both for $\Sigma$-types and to denote (presentations of) gats. It should be clear from the context which one is meant.)

In Section \ref{level-indexed-tt} we will present $\Sigmaint$, the finitary gat of level-indexed cwfs with extra structure for the small type formers of Martin-Löf type theory and for level-indexed universes. An indexed cwf consists of a base category $\C$ and a cwf-valued presheaf
$$
P : \C^\op \to \CwF
$$
We get a model for universe polymorphism by letting $\C$ be the category of contexts of a unityped cwf (ucwf) of levels and by equipping $\CwF$ with extra structure for the small type formers. We then
add extra structure for level-indexed universes and level-indexed products of types.

A presentation of type theory by a gat of cwfs is a higher-level
%\footnote{PD: removed "and more objective"}
notion than a presentation by grammar and inference rules. This is because the gat only records the important rules and highlights categorical structure. When building initial models in terms of grammar and inference rules, we need to make various syntactic choices and to include numerous bookkeeping rules, such as general rules of equality. There will be a multitude of options. We hope that our high level of abstraction will allow to prove equivalence between different options and claim that there is indeed a unique abstract notion of the mathematical theory under consideration.

The present work can be viewed more broadly as a case study for categorical logic based on gats and cwfs. The idea to replace traditional deductive systems by more regular mathematical notions goes back to the early days of categorical logic, as envisaged by Lawvere \cite{lawvere:tac-adjointness}:
\begin{quotation}
\emph{
My 1963 observation (referred to by Eilenberg and Kelly in La Jolla, 1965), that cartesian closed categories serve as a common abstraction of type theory and propositional
logic, permits an invariant algebraic treatment of the essential problem of proof theory,
though most of the later work by proof theorists still relies on presentation-dependent
formulations. ...
%The strategy to interpret proofs themselves as structures had
%been discussed by Kreisel; however, the influential “realizers” of Kleene are not yet the
%usual mathematical sort of structures.
}
\end{quotation}
A similar view was expressed by Voevodsky \cite{voevodsky:cmu2010}:
\begin{quotation}
\emph{
I will speak about type systems. It is diﬃcult for a mathematician since
a type system is not a mathematical notion. I will spend a little time
explaining how I see ``type systems” mathematically. ...}

\emph{Thesis 0. Any formal deduction system can be specified in the form of a
quasi-equational theory.} ...

\emph{Fact 1. Any quasi-equational theory has an initial model.} ...

\emph{This view of formal deduction systems has many advantages. One is that it
suggests a uniform approach to the formal description of various deductive
systems. Another one is that ”interpretations” of the deductive system are
directly connected with the models of the corresponding quasi-equational
theory.}
\end{quotation}
\ERASER{\todo[inline]{Already here we could say: we pursue a variant of Voevodsky's approach: we will use gats rather than quasi-equational theories, and we will use cwfs rather than the C-systems (contextual categories). Gats and cwfs are intimately connected ... cwfs (and variants) are suitable for a uniform approach to deductive logical systems.
After mentioning the initiality conjecture project we could mention our generic construction of an initial model of a gat.}}

Voevodsky considered these issues important for the development of his Univalent Foundations of Mathematics. To advance the state
of the art he proposed the {\em Initiality Conjecture project} the goal of which is to define a general class of dependent type theories and to develop generic metatheory for theories in this class. We quote from the introduction of an extended abstract where Voevodsky \cite{voevodsky:initiality} motivates his project:
\begin{quotation}
\emph{The first few steps in all approaches to the set-theoretic semantics of dependent
type theories remain insuﬃciently understood. The constructions which have been
worked out in detail in the case of a few particular type systems by dedicated authors
are being extended to the wide variety of type systems under consideration today by
analogy. This is not acceptable in mathematics. Instead we should be able to obtain
the required results for new type systems by specialization of general theorems and
constructions formulated for abstract objects the instances of which combine together
to produce a given type system.}
$$
\vdots
$$
\emph{A crucial component of this approach is the expected result that for a particular
class of inference rules the term model is an initial object in the category of models.
This is known as the Initiality Conjecture. In the case of the pure Calculus of
Constructions with a “decorated” application operation this conjecture was proved in
1988 by Thomas Streicher \cite{streicher:thesis}. The problem of finding an appropriate formulation
of the general version of the conjecture and of proving this general version will be the
subject of future work.}
\end{quotation}
On the surface such initiality proofs may seem straightforward, but they depend on subtle details in the formulation of grammar and inference rules. This is why Voevodsky insisted on calling such theorems ``conjectures'' until proven rigorously and, ideally, implemented in a proof assistant.

An example of an implemented initiality proof is Brunerie and de Boer's \cite{Brunerie:initiality,deBoer:lic} proof in Agda that a version of Martin-Löf type theory with an external tower of universes is an initial contextual category \cite{cartmell:phd,cartmell:apal} with appropriate extra structure. %Theversion of type theory with de Bruijn variables and implicit substitution.

In this article we propose an approach to Voevodsky's project based on gats and cwfs. As shown in our article \cite{bezem:hofmann} on gats and cwfs, for each finite presentation $\Sigma$ of a gat, there is a general construction of a term model $\T_\Sigma$ and this is initial in the category $\CwF_\Sigma$ of categories with families (cwfs) with a $\Sigma$-structure. The idea is to capture a logical theory $\T$ by a corresponding gat $\Sigma$ such that (the term model of) $\T$ is isomorphic to (the externalization of) $\T_\Sigma$. Externalization is explained in Section \ref{initiality-external}.

Gats and cwfs are intimately connected. Cwfs can themselves be presented as a gat $\Sigma^\CwF$, and we can extend this gat with operator symbols and equations for the type formers of Martin-Löf type theory with an external tower of universe yielding the gat $\Sigmaext$. Although this gat is infinitary, it has an initial model $\T_{\Sigmaext}$  obtained by extending the general construction of an initial model for finitary gats~\cite[Section 5.7]{bezem:hofmann}.

This sums up our approach to Voevodsky's project. We characterize dependent type theories as initial models of gats of cwfs with extra structure or, as we shall see below, gats of variations of cwfs. In this way we have generic constructions of term models but can also study alternative constructions of initial models. The reason for focusing on gats of cwfs (with extra structure) is that they occupy an intermediate place between dependent type theories defined by grammar and inference rules and notions of model based on more mainstream categorical constructions. In particular, the gat of cwfs resembles Martin-Löf's substitution calculus for dependent type theory \cite{martinlof:gbg92,tasistro:lic}.

Our approach to Voevodsky's project extends beyond dependent type theory. For example, by considering simply typed cwfs (scwfs) and unityped cwfs (ucwfs) we can also capture various simply typed and untyped logical systems as gats and thus widening the scope of {\em uniform categorical logic} based on gats and cwfs, see Castellan, Clairambault, and Dybjer \cite{castellan:lambek}.

Moreover, variations of indexed cwfs capture various other logical systems and can be formalized as gats. For example, untyped predicate logic can be captured by ucwf-indexed scwfs with extra structure for the logical constants. Typed predicate logic can be captured by scwf-indexed scwfs and dependently typed predicate logic (Makkai \cite{makkai:folds}, Gambino and Aczel \cite{gambino-aczel}, Belo \cite{belo}, and Palmgren \cite{Palmgren19}) by cwf-indexed scwfs, both with suitable extra structure for type formers and logical constants.

\paragraph{Universe polymorphism.} We refer to our paper \cite{BezemCDE22} for a presentation of the inference rules of Martin-Löf type theory with explicit universe polymorphism, where the reader can also find motivation, examples, and a discussion of related work. Here we only give a brief overview.

An implicit form of universe polymorphism was introduced by Huet \cite{Huet87} and is an essential feature of the proof assistant Rocq (Coq) \cite{rocq:general}. Alternatively, Agda \cite{agda-wiki} and Lean \cite{moura:lean} employ versions of universe polymorphism where universe levels are declared explicitly.

As already mentioned, we followed Courant's approach \cite{Courant02} with universe level judgments. We also presented an extension where equational constraints between universe levels can be declared (\cite[Section 5]{BezemCDE22}), building on a proposal by Voevodsky \cite{VV}. %This extension can also be described by a gat, but we postpone this topic to a forthcoming paper.

\paragraph{Dedication.} We dedicate this article to professor Stefano Berardi, the University of Torino. Stefano is a friend, colleague, and coauthor, who spent the winter and spring 1993/94 in the type theory group in Göteborg. He has made fundamental contributions to type theory and constructivity and in particular to the constructive content of classical logic and the constructive analysis of impredicativity.

\section{Type theory with an external tower of universes}\label{sigmaext}

\subsection{The gat of cwfs} A cwf consists of a category $\C$ of contexts and context morphisms (substitutions) with a terminal object, a family-valued presheaf
$$
T : \C^\op \to \Fam
$$
and a notion of context comprehension. Here $\Fam$ is the category of indexed families of sets $(A,B)$, where $A$ is an index set and $B$ is a family of sets indexed by $A$. The object part of the presheaf $T$ maps a context to a family of terms indexed by a type. The arrow part represents substitution in types and terms. We often refer to a cwf as a quadruple $(\Ctx,\Sub,\Ty,\Tm)$, where $(\Ctx,\Sub)$ are the objects and arrows of the category of contexts $\C$, and $(\Ty,\Tm)$ refer to the family $T$ of terms indexed by types, both in context. The reader is referred to Dybjer \cite{dybjer:torino}, Hofmann \cite{hofmann:cambridge}, and Castellan et al.\ \cite{castellan:lambek} for a full definition and further information about cwfs.

The gat of cwfs with extra structure for $\Pi$-types, a notion of model of Martin-Löf type theory with $\Pi$-types, was presented by Dybjer \cite{dybjer:torino}. The extra operator symbols and equations needed for the natural number type and one universe can be found in our article \cite{bezem:hofmann}. We will use the same notation for gats as in the latter paper.

\ERASER{\todo[inline]{Add something about the general correspondence between gats and type theory: sort symbol = judgment form, operator symbol for formation, introduction, and elimination rules, equation for equality rules.}}

\paragraph{Sort symbols.}
The gat of cwfs has four sort symbols:
\begin{eqnarray*}
&\vdash& \ctx\\
\Delta, \Gamma : \ctx&\vdash& \sub(\Delta,\Gamma)\\
\Gamma : \ctx&\vdash& \ty(\Gamma)\\
\Gamma : \ctx, A : \ty(\Gamma)&\vdash& \tm(\Gamma,A)
\end{eqnarray*}
corresponding to the objects and morphisms of the category of contexts, and the family of terms indexed by types in a given context, respectively. Thus $\ctx$ is a constant sort symbol, $\sub$ is a binary sort symbol depending on arguments $\Delta, \Gamma : \ctx$, $\ty$ is a unary sort symbol depending on the argument $\Gamma : \ctx$, and $\tm$ is a binary sort symbol depending on arguments $\Gamma : \ctx, A : \ty(\Gamma)$.

\paragraph{Operator symbols.}
\ERASER{The gat of cwfs has operator symbols for the basic operations of cwfs (identity and composition of context morphisms, substitution in types and terms, empty context and context morphism, context extension and context morphism extension, projection morphism, last variable term)}
\begin{eqnarray*}
\Gamma : \ctx &\vdash& \id_{\Gamma} : \sub(\Gamma,\Gamma)\\
\Xi,\Delta,\Gamma : \ctx, \gamma : \sub(\Delta,\Gamma), \delta : \sub(\Xi,\Delta) &\vdash&
\gamma \circ \delta : \sub(\Xi,\Gamma)\\
\\
\Gamma,\Delta : \ctx, A:\ty(\Gamma), \gamma : \sub(\Delta,\Gamma) &\vdash&
A[\gamma] : \ty(\Delta)\\
\Gamma,\Delta : \ctx, A:\ty(\Gamma), \gamma : \sub(\Delta,\Gamma), a:\tm(\Gamma,A) &\vdash&  a[\gamma] : \tm(\Delta,A[\gamma])\\
\\
&\vdash& 1 : \ctx\\
\Gamma : \ctx &\vdash& \tuple{}_\Gamma : \sub(\Gamma,1)\\
\\
\Gamma : \ctx, A:\ty(\Gamma) &\vdash& \Gamma\cext A : \ctx\\
\Gamma,\Delta : \ctx, A:\ty(\Gamma), \gamma : \sub(\Delta,\Gamma), a:\tm(\Delta,A[\gamma]) &\vdash& \tuple{\gamma,a} : \sub(\Delta,\Gamma\cext A)\\
\Gamma : \ctx, A:\ty(\Gamma) &\vdash& \p_{\Gamma,A}: \sub(\Gamma\cext A,\Gamma)\\
\Gamma : \ctx, A:\ty(\Gamma) &\vdash& \q_{\Gamma,A}: \tm(\Gamma\cext A,A[\p])
\end{eqnarray*}
The first line states that identity is a unary operator symbol with argument $\Gamma : \ctx$. The second line states that $\circ$ is a binary operator symbol with five arguments $\Xi,\Delta,\Gamma : \ctx, \gamma : \sub(\Delta,\Gamma), \delta : \sub(\Xi,\Delta)$ and result sort $\sub(\Xi,\Gamma)$. Note that only two of the five official arguments are explicit. To alleviate notation, we often suppress arguments of operator symbols. Note also that we overload notation for type and term substitution $A[\gamma]$ and $a[\gamma]$. Moreover, we sometimes drop further arguments and write $\id, \tuple{},\p,\q$ without the arguments in index position.

\paragraph{Equations.}
The gat of cwfs has 13 equations. We illustrate the notation by showing one of the laws for identity morphisms:
\begin{eqnarray*}
\Delta, \Gamma : \Obj, \gamma : \sub(\Delta,\Gamma) &\vdash& \id_\Gamma \circ \gamma = \gamma : \sub(\Delta,\Gamma)
\end{eqnarray*}
Moreover, we often drop argument and result types in equations:
\begin{eqnarray*}
\id_\Gamma \circ \gamma = \gamma
\end{eqnarray*}
Note that there are specific cases where it is essential to keep the context explicit. Assume that we add an operator symbol for the identity type former $\Id$ in Martin-Löf type theory:
\begin{eqnarray*}
\Gamma : \ctx, A : \ty(\Gamma), a, a' : \tm(\Gamma, A) &\vdash \Id_{\Gamma,A}(a,a'): \ty(\Gamma)
\end{eqnarray*}
In extensional Martin-Löf type theory \cite{martinlof:hannover} we have the rule of equality reflection. This can be captured by an equation in gats:
\begin{eqnarray*}
\Gamma : \ctx, A : \ty(\Gamma), a, a' : \tm(\Gamma, A),p :  \tm(\Gamma,\Id_{\Gamma,A}(a,a'))&\vdash& a = a' : \tm(\Gamma,A)
\end{eqnarray*}
Note that the variable $p$ does not occur on the right hand side.
However, it is an important variable expressing
the condition under which $a=a'$ holds and must not be omitted.

We refer to \cite{bezem:hofmann} for the remaining cwf-equations. The reader may also consult Appendix A where the equations for level-indexed cwfs are displayed.

\paragraph{Interpretation in cwfs.}
A model of a gat $\Sigma$ is an object in the category $\CwF_\Sigma$ of cwfs with extra structure for the sort symbols and operator symbols in $\Sigma$ satisfying the equations \cite{bezem:hofmann}. Sort symbols are interpreted as cwf-types and operator symbols are interpreted as cwf-terms, both in context. For example, in the term model cwf of the gat for cwfs $\Sigma^\CwF$ we have $\ctx \in \Ty(1), \sub \in \Ty(1.\ctx.\ctx[\p])$. Moreover, as an example of an operator symbol, $\id \in \Tm(1.\ctx,\sub(\q,\q))$ where $\sub(\Delta,\Gamma)$ is shorthand for
$\sub[\tuple{\Delta,\Gamma}]$, $\tuple{a,b}$ abbreviates $\tuple{\tuple{a}, b}$, $\tuple{a}$ abbreviates $\tuple{\tuple{}, a}$, etc. We use $\id_\Gamma$ is shorthand for $\id[\tuple{\Gamma}]$, etc. An object in $\CwF_{\Sigma^\CwF}$ is a cwf with an internal cwf , see \cite{bezem:hofmann} for more information.

\subsection{The gat of cwfs with an external tower of universes}
In \cite[Section 3]{BezemCDE22} we displayed the inference rules for Martin-Löf type theory with an external tower of universes $\UU_l$, where $l \in \Nbb$ is an external natural number. Here we show the operator symbols and equations for the corresponding gat.

We assume that we already have defined the gat of cwfs with the extra structure for the standard small type formers $\Pi, \Sigma, \N_0, \N_1, \N_2, \N, \W$, and $\Id$, and we wish to define a tower of universes closed under those. To save space, we shall only display the operator symbols and equations for closure under $\Pi$-types, since it is straightforward to add similar operator symbols for closure under the other small type formers. We shall use the same convention throughout the paper.

\paragraph{Operator symbols.}
The operator symbol for $\Pi$-formation is
\begin{eqnarray*}
\Gamma : \ctx, A : \ty(\Gamma), B : \ty(\Gamma.A) &\vdash& \Pi(A,B) : \ty(\Gamma)
\end{eqnarray*}
and we refer to \cite{bezem:hofmann} for the operator symbols for abstraction and application, equations for the $\beta$ and $\eta$ rule, and equations expressing that $\Pi$, abstraction, and application commute with term substitution. We also refer to Appendix A for the level-indexed version.

In the gat for externally indexed universes we have the following families of operator symbols (the universes $\UU_l$, the decoding maps $\Ta_l$, codes for $\Pi$, and codes $\UU_l^m$ for $\UU_l$ in $\UU_m$ for $l, l' , m \in \Nbb$ with $l < m$):
\begin{eqnarray*}
\Gamma : \ctx &\vdash& (\U_{l})_\Gamma : \ty(\Gamma)\\
\Gamma : \ctx, a : \tm(\Gamma,(\U_{l})_\Gamma) &\vdash& {\Ta_{l}}(a) : \ty(\Gamma)\\
%\Gamma : \ctx &\vdash& (\N^0)_\Gamma : \tm(\Gamma,(\U_0)_\Gamma) \\
\Gamma : \ctx,
a : \tm(\Gamma,(\U_{l})_\Gamma),
b :  \tm(\Gamma . \Ta_{l}(a), (\U_{l'})_\Gamma)
&\vdash&
 \Pihat^{l,l'}(a,b) : \tm(\Gamma,(\U_{l \vee l'})_\Gamma)\\
 \Gamma : \ctx&\vdash&(\UU^m_l)_\Gamma: \tm(\Gamma,(\UU_{m})_\Gamma)
\end{eqnarray*}
Note that codes for $\Pi$ are doubly indexed and $l \vee l' = \max(l,l')$.
We have again left some arguments to operator symbols implicit. For example, the decoding operators $\Ta_l$ are binary operators with official notation $\Ta_l(\Gamma,a)$, but above we omitted $\Gamma$ and wrote $\Ta_l(a)$.

\paragraph{Equations.}
We have the following decoding equations:
\begin{eqnarray*}
%\Ta(\N^0_\Gamma) &=& \N_\Gamma\\
\Ta_{l \vee l'}(\Pi^{l,l'}(a,b)) &=& \Pi(\Ta_l(a),\Ta_{l'}(b))\\
\Ta_{m}((\UU^m_l)_\Gamma) &=& (\UU_l)_\Gamma
\end{eqnarray*}
The operator symbols commute with term substitution:
\begin{eqnarray*}
(\U_l)_\Gamma [ \gamma ] &=& (\U_l)_\Delta\\
\Ta_l(a) [ \gamma ] &=& \Ta_l(a[ \gamma ] )\\
%\N^0_\Gamma [ \gamma ] &=&\N^0_\Delta\\
\Pi^{l,l'}(a,b)[ \gamma ] &=& \Pi^{l,l'}(a [ \gamma ], b[ \gamma^\dagger ])\\
(\UU^m_l)_\Gamma[ \gamma ] &=&(\UU^m_l)_\Delta\\
\Ta^m_{l}(a)[\gamma] &=& \Ta^m_{l}(a[\gamma])
\end{eqnarray*}
where $\gamma : \sub(\Delta,\Gamma)$ and $\gamma^\dagger = \langle \gamma \circ \p_{\Delta,\Ta_l(a[\gamma])}, \q_{\Delta,\Ta_l(a[\gamma])}\rangle : \sub(\Delta.\Ta_l(a[\gamma]),\Gamma.\Ta_l(a))$.
%\paragraph{Adding cumulativity.}

If we want a cumulative tower of universes we add operator symbols that lift elements in the $l$th universe to the $m$th universe for $l < m$:
\begin{eqnarray*}
\Gamma : \ctx, a : \tm(\Gamma,(\U_{l})_\Gamma) &\vdash& {\Ta^m_{l}}(a) : \tm(\Gamma,(\U_{m})_\Gamma))
\end{eqnarray*}
with the decoding
\begin{eqnarray*}
\Ta_m(\Ta^m_{l}(a)) &=& \Ta_l(a)
\end{eqnarray*}
It commutes with substitution:
\begin{eqnarray*}
\Ta^m_{l}(a)[\gamma] &=& \Ta^m_{l}(a[\gamma])
\end{eqnarray*}

In the presence of cumulativity it suffices that codes for $\Pi$ have one superscript rather than two. However, we
 do not display this simplification here but refer to section  \ref{cumulativity} on cumulativity for level-indexed universes.

This concludes the presentation $\Sigmaext$ of a gat for cwfs with small type formers and an external tower of universes.

\subsection{Syntax and inference rules as an initial model}\label{initiality-external}

\paragraph{A generic construction based on explicit substitution.} In our article  \cite{bezem:hofmann} on gats and cwfs we defined the notion of a correct {\em presentation} $\Sigma$ (a finite list of sort symbols, operator symbols, and equations) of a gat and the associated category of models $\CwF_\Sigma$ of cwfs with a $\Sigma$-structure. We then constructed for each $\Sigma$ an initial object $\T_\Sigma$ in $\CwF_\Sigma$ as a type theory defined in terms of a grammar and inference rules. This type theory is a calculus of explicit substitution, where raw expressions are formed by untyped cwf-combinators and untyped versions of the operator symbols. See Abadi, Cardelli, Curien and L\`evy \cite{AbadiCCL90} for an explicit substituion calculus in a simply typed setting.

In the current section we have presented the {\em infinitary} gat $\Sigmaext$ of cwfs with extra structure for the small type formers and a tower of universes. However, the above construction of an initial model applies to {\em finitary} gats. Nevertheless, as explained in \cite[Section 5.7]{bezem:hofmann}, we can generalize our construction to some non-finitely presented gats. If we have an increasing sequence of finite presentations $\Sigma_n$ we can build the initial model $\T_\Sigma$ of their union $\Sigma$ in stages. In this way we can build an initial model of $\Sigmaext$ as a union of $\Sigma_n$ -- the gats of cwfs with extra structure for the small type formers and a truncated tower of $n$ universes.

Furthermore, $\CwF_{\Sigmaext}$ is the category of cwfs with an {\em internal} cwf with extra structure for the small type formers and a tower of universes. In contrast to this we have the category of cwfs with extra structure for the small type formers and a tower of universes $\CwFext$. A cwf with an internal cwf $(\Ctx,\Sub,\Ty,\Tm,\ctx,\sub,\ty,\tm)$ in $\CwF_{\Sigmaext}$ determines an {\em external} cwf $(\Ctx^*,\Sub^*,\Ty^*,\Tm^*)$ in $\CwFext$ with extra structure for the small type formers and a tower of universe  as follows:
\begin{eqnarray*}
\Ctx^* &=& \Tm(1,\ctx)\\
\Sub^*(\Delta,\Gamma) &=& \Tm(1,\hom(\Delta,\Gamma))\\
\Ty^*(\Gamma) &=& \Tm(1,\ty(\Gamma))\\
\Tm^*(\Gamma,A) &=& \Tm(1,\tm(\Gamma,A))
\end{eqnarray*}
%Note that $\Delta,\Gamma \in \Tm(1,\ctx)$ are internal contexts, and not external ones $\Delta,\Gamma \in \Ctx$. We leave the verification of this construction to future work.
%\footnote{PD: Thierry and I discussed the externalization process before we submitted the paper in August. The verification doesn't seem difficult, but relies on the definitions in the paper about gats and cwfs. We decided not to include it.
%
%The operator symbols become
%\begin{eqnarray*}
%\id^*_\Gamma &=& \id_1[{\tuple{\Gamma}}]\\
%\gamma \circ^* \delta &=&\circ[ \tuple{\gamma,\delta[\p]?} ]\\
%&\vdots&
%\end{eqnarray*}
%Case of monoids: one sort symbol and two operator symbols:
%\begin{eqnarray*}
%&\vdash& m\\
%&\vdash& e : m\\
%a, b : m &\vdash& a + b : m
%\end{eqnarray*}
%In the initial (?) cwf with an internal monoid:
%\begin{eqnarray*}
%m &\in& \Ty(1)\\
%e &\in& \Tm(1,m)\\
%+ &\in& \Tm(1.m.m[\p],m[p^2])
%\end{eqnarray*}
%The externalized version. We get the external monoid $(m^*,e^*,+^*)$:
%\begin{eqnarray*}
%m^* &=&\Tm(1,m)\\
%e^* &=& e\\
%+^*(a, b) &=& +[ \tuple{a,b} ]
%\end{eqnarray*}
%Note that $a,b \in \Tm(1,m)$ while $a,b$ in $a, b : m \vdash a + b : m$ are named variables for a readable presentation of the typing of an operator symbol with interpretation $+ \in \Tm(1.m.m[\p],m)$ in the cwf.}

\paragraph{A construction based on implicit substitution (initiality conjecture).} We contrast type theories with {\em implicit substitution}, where substitution is defined by structural induction, with type theories with {\em explicit substitution}, where substitution is a syntactic constructor of expressions.
We shall outline an alternative construction of an initial object in $\CwFext$ based on $\TText$ -- the type theory with implicit substitution and an external tower of universes presented in \cite[Section 3]{BezemCDE22}.

As already mentioned we use decorated expressions; for example, we decorate application with type information. We refer to this as the {\em raw syntax}, that is, expressions that are not necessary well-typed. In the raw syntax we include raw context morphisms, although such are not mentioned in loc.cit. These are lists of raw terms, where $\tuple{}$ denotes the empty list, and $\tuple{\gamma,a}$ denotes the list $\gamma$ extended by a new term $a$.

We build an object $(\Ctx,\Sub,\Ty,\Tm)$ of $\CwFext$ based on the type theory with implicit substitution $\TText$ as follows. First, we interpret the sort symbols in terms of the judgment forms of $\TText$ as follows:
\begin{itemize}
\item $\Gamma \in \Ctx$ is defined as $\Gamma \vdash$ quotiented by the equivalence relation of context equality $\Gamma = \Gamma' \vdash$. The latter is not stated explicitly in loc.cit. but can easily be added.
\item $A \in \Ty(\Gamma)$ is defined as $\Gamma \vdash A$ quotiented by the equivalence relation $\Gamma \vdash A = A'$.
\item $a \in \Tm(\Gamma,A)$ is defined as $\Gamma \vdash a : A$ quotiented by the equivalence relation $\Gamma \vdash a = a' : A$.
\item There are no explicit judgments $\Delta \vdash \gamma : \Gamma$ and $\Delta \vdash \gamma = \gamma' : \Gamma$  in loc.cit, but these can be defined in terms of $\Delta \vdash a : A$ and $\Delta \vdash a = a' : A$. Then $\gamma \in \Sub(\Delta,\Gamma)$ can be defined as $\Delta \vdash \gamma : \Gamma$ quotiented by the equivalence relation $\Delta \vdash  \gamma = \gamma' : \Gamma$.
\end{itemize}
We then define the operator symbols on equivalence classes. First there are the operator symbols that construct raw syntax: the basic cwf-combinators $1, -.-, \tuple{}, \tuple{-,-}$,  $\Pi, \lambda, \app$ for $\Pi$-types, and similarly for the other small type formers, and $\U_l, \Ta_l, \U^m_l$ and $\Ta^m_l$ for the tower of universes. We just give two examples.
\begin{itemize}
\item The empty context $1$ is a constructor of raw contexts and we define the terminal object in the term model as the equivalence class of the empty context $[1] \in \Ctx$.
\item Context extension is also a constructor of raw contexts. We define the extension of $[\Gamma] \in \Ctx$ with $[A] \in \Ty([\Gamma])$ as $[\Gamma.A] \in \Ctx$ and show that this operation respects the equivalence,
so that it extends to the equivalence classes.
%\item Similarly for the constructors of raw cwf-morphisms $\tuple{}$ and $\tuple{-,-}$.
%\item The operator symbols for the type formers, such as $\Pi, \lambda, \app$ for $\Pi$-types and $\U_l, \Ta_l, \U^m_l$ and $\Ta^m_l$ are also constructors of raw syntax.
\end{itemize}
Then we consider the operator symbols that correspond to operations defined by induction on the structure of raw expressions. We first define the substitution in types and terms in the model:
\begin{itemize}
\item If $A$ is a raw type and $\gamma$ is a raw substitution, we first define the result $A[\gamma]$ of substituting $\gamma$ in $A$ by induction on the structure of $A$. We need to show that this operation preserves equivalence classes.
\item Substitution in raw terms is defined similarly.
%if $a$ is a raw term and $\gamma$ is a raw substitution, we first define the result $a[\gamma]$ of substituting $\gamma$ in $a$ by induction on the structure of the latter. Then we raise this operation to equivalence classes by proving that equality of terms is preserved.
\item We can also define the other implicit operations $\id, \circ, \p, \q$.
\end{itemize}
To prove that we construct an object in $\CwFext$, we need to check the gat-equations. Then we need to show that there is a unique morphism in $\CwFext$ to any other object.

We can compare the initiality proof outlined above and the proof implemented in Agda by Brunerie and de Boer \cite{Brunerie:initiality,deBoer:lic}, since their proof is of a version of type theory with an external tower of universes similar to ours. One difference is that their universes are \`a la Russell, while we present both versions \`a la Tarski and \`a la Russell. Moreover, they just have the rule $\UU_l : \UU_{l+1}$ and only consider the non-cumulative case, while we have $\UU_l^m$ for $\UU_l$ in any larger universe $\UU_m$ and consider both the non-cumulative and cumulative cases. On the categorical side, there is the relatively minor difference between contextual categories and cwfs, where we note that initial cwfs (with extra structure) are contextual \cite{ClairambaultD14,castellan:lambek}.
%However, their notion of model is not expressed explicitly as a gat.

\section{Level-indexed type theory}\label{level-indexed-tt}

\subsection{The ucwf of levels}\label{ucwf-levels}
As already mentioned in the introduction, in our paper \cite{BezemCDE22} we added universe level {\em judgments}
$$
l\ \level
\hspace{5em}
l = l'
$$
to the usual judgment forms of Martin-Löf type theory. Moreover, universe level variables and ordinary term variables can be declared in any order. To simplify the correspondence with the gat formalization, we will in the sequel assume that contexts have the form $n, \Gamma$, where $n$ is the number of (de Bruijn) level variables, and $\Gamma$ is an ordinary context that depends on these $n$ level variables.
We have a next level function $(-)^+$ and an operation $\vee$ for join of levels. However, as in \cite[Section 3,4]{BezemCDE22}, we have no level 0 for the first universe. It follows that all universes are polymorphic.
Levels form an upper semilattice with respect to $\vee$, and $(-)^+$ commutes
with $\vee$ and is inflationary, see below.

\paragraph{Sort symbols.} Levels can be organized as a ucwf. We have the following sort symbols:
\begin{eqnarray*}
&\vdash& \lctx\\
m, n : \lctx &\vdash& \lHom(m,n)\\
m : \lctx &\vdash& \ltm(m)
\end{eqnarray*}
standing for level context, level context morphism (substitution), and level term.
Since ucwfs are cwfs with only one type, we do not need a sort symbol for level types.

\paragraph{Operator symbols.} The operator symbols for ucwfs are simplified versions of those for cwfs, where all dependence on types is removed:
%\footnote{We use  $\sigma,\tau$ for level substitutions.}
\begin{eqnarray*}
m : \lctx &\vdash& \lid_m : \lhom(m,m)\\
m, n, p : \lctx, \sigma : \lhom(n,p), \tau : \lhom(m,n) &\vdash&
\sigma \circ \tau : \lhom(m,p)\\
\\
m,n: \lctx, \sigma : \lhom(n,m), l :\ltm(m) &\vdash&  l[\sigma] : \ltm(n)\\
\\
&\vdash& 0 : \lctx\\
m : \lctx &\vdash& \tuple{}_m : \lhom(m,0)\\
\\
m : \lctx &\vdash& \s(m) : \lctx\\
m,n : \lctx, \sigma : \lhom(n,m), l:\ltm(n) &\vdash& \tuple{\sigma,l} : \lhom(n,\s(m))\\
m : \lctx &\vdash& \lp_m: \lhom(\s(m),m)\\
m : \lctx &\vdash& \lq_m: \ltm(\s(m))
\end{eqnarray*}
Note that we no longer need an operator symbol for substitution in types but only in level terms. We also change the notation to suggest that this is an operation on levels. For example, we use $\sigma$ and $\tau$ to range over level substitutions to distinguish them from term substitutions $\gamma$ and $\delta$. However, we keep the notation $\circ$ for composition of level substitutions, $l[\sigma]$ for level substitution in level terms, $\tuple{}_m$ for the empty level context morphism and $\tuple{\sigma,l}$ for level context morphism extension. The notation for level contexts suggests that we have an initial ucwf where $n : \lctx$ is a natural number that records the number of available level variables. Thus $0 : \lctx$ is the terminal object in the ucwf.

The ucwf of levels also has operator symbols for next level and join of two levels:
\begin{eqnarray*}
m : \lctx, l : \ltm(m) &\vdash& l^+ : \ltm(m)\\
m : \lctx, l,l' : \ltm(m) &\vdash& l \vee l' : \ltm(m)
\end{eqnarray*}
%\end{tiny}

%\begin{tiny}
\paragraph{Equations.}
The ucwf-equations are the cwf-equations (see Dybjer \cite{dybjer:torino} and our joint paper \cite{BezemCDE22})
%appendix   \ref{sec:gatPiU})
for the special case that there is only one type, so that all type equations are redundant:
\begin{eqnarray*}
\lid_{n} \circ \sigma &=& \sigma \\
 \sigma \circ \lid_{n} &=& \sigma \\
(\sigma \circ \tau) \circ \upsilon &=& \sigma \circ (\tau \circ \upsilon)\\
%&\vdash&
%A[\id_{n}] &=& A \\
%a:\tm_n(\sigma,A) &\vdash&
l[\lid_{n}] &=& l \\
%n : \lctx, \Xi,\Delta,\Gamma : \ctx_n, \delta : \sub_n(\Xi,\Delta), \gamma : \sub_n(\Delta,\Gamma),
%&\vdash&
%A[\gamma\circ\delta] &=&A[\gamma][\delta]\\
%: \ty_n(\Xi)\\
%n : \lctx, \Xi,\Delta, : \ctx_n, \delta : \sub_n(\Xi,\Delta), \gamma : \sub_n(\Delta,\Gamma),
%a:\tm_n(\Gamma,A) &\vdash&
l[\sigma\circ\tau] &=& l[\sigma][\tau]\\
%: \tm_n(\Xi,A[\gamma\circ\delta])\\
%n : \lctx &\vdash&
\lid_{0} &=& \tuple{}_{0}\\
 %: \sub_n(1_n,1_n)\\
%\Gamma,\Delta : \ctx_n , \gamma : \sub_n (\Delta,\Gamma) &\vdash&
\tuple{}_{n}\circ\sigma &=& \tuple{}_{m}\\
\lp_n \circ \tuple{\sigma,l} &=& \sigma\\
\lq_n [\tuple{\sigma,l}] &=& l\\
\tuple{\sigma,l} \circ \tau &=& \tuple{\sigma \circ \tau,l[\tau]}\\
\lid_{\s(n)} &=& \tuple{\lp_n,\lq_n}
\end{eqnarray*}
The semi-lattice equations for $l \vee l'$ are:
\begin{eqnarray*}
(l \vee l') \vee l'' &=& l \vee (l' \vee l'')\\
l \vee l' &=& l'\vee l\\
l \vee l &=& l
\end{eqnarray*}
and the equations for the inflationary endofunction $(-)^+$ are:
\begin{eqnarray*}
l \vee l^+ &=& l^+\\
(l\vee l')^+ &=& l^+\vee l'^+
\end{eqnarray*}
The operator symbols $\vee$ and $+$ commute with level substitution:
\begin{eqnarray*}
(l \vee l')[\sigma] &=& l[\sigma] \vee l' [\sigma]\\
 l^+[\sigma] &=&  l[\sigma]^+
\end{eqnarray*}
%\end{tiny}
\paragraph{Lawvere theories.} We remark that ucwfs are similar to Lawvere theories, but are closer to the usual syntax based on $n$-place functions. One can prove that Lawvere theories are equivalent to contextual ucwfs, that is, ucwfs where each context has a length \cite{ClairambaultD14,castellan:lambek}.

\paragraph{Level equality sorts.}
When we encode type theory in gats, the principle is to introduce one sort for each main form of judgment. For example, $l : \ltm(n)$ represents the judgment $n \vdash l\ \level$. Equality judgments are then represented by equalities: $l = l' : \ltm(n)$ represents $n \vdash l = l'$.

As we shall see in the next subsection, when typing the codes $\UU_l^m$ for universes $\UU_l$ in $\UU_m$ we need to express the constraint that $l < m$ which is defined as $l^+ \vee m = m$. However, equalities are not allowed as assumptions in gats. Therefore, we add a new sort symbol for level equality:
\begin{eqnarray*}
n : \lctx, l, l'  : \ltm(n) &\vdash& \leq_n(l,l')
\end{eqnarray*}
and an operator symbol for reflexivity:
\begin{eqnarray*}
n : \lctx, l : \ltm(n) &\vdash& \refl(l) :  \leq_n(l,l)
\end{eqnarray*}
Now we can express the constraint $l < m$ by assuming $p : \leq_n(l^+ \vee m, m)$. Note that if $l =  l' : \ltm(n)$ in the term model (see section \ref{initiality-internal}), that is, if $l = l'$ can be derived by equational reasoning from the laws for $\vee$ and $(-)^+$, then $\refl(l) :  \leq_n(l,l')$
%\begin{eqnarray*}
%n : \lctx, l, l'  : \ltm(n) &\vdash& l = l' : \ltm(n)
%\end{eqnarray*}
%in the initial model, that is, if $l = l'$ can be derived by equational reasoning from the laws for $\vee$ and $(-)^+$ , then we can also derive
%\begin{eqnarray*}
%n : \lctx, l,l' : \ltm(n) &\vdash& \refl(l) :  \leq_n(l,l')
%\end{eqnarray*}
by preservation of equality, a principle available in all gats. In the opposite direction we have the following:
%\footnote{Discussion between Peter and Marc 250930: We should have the construction of the initial ucwf of levels already here. We also need to immediately interpret level equality sorts. Then we have the means to prove the following proposition.
%We should discuss both the model with quotients and the model of normal forms.}
\begin{proposition*}
If $n : \lctx, l, l' : \ltm(n),$ and $p :  \leq_n(l,l')$ in the initial model,
then $p = \refl(l) : \leq(l,l')$ and $l = l' : \ltm(n)$.
\end{proposition*}

This can be proved by a normal form argument. As remarked by Bezem and Coquand \cite{bezem-coquand:lattices},
each level term has a normal form
$
\alpha_1^{+^{p_1}} \vee \cdots \vee \alpha_m^{+^{p_m}}
$
for $p_i \geq 0$ for $1 \le i \le m \le n$ and level variables (de Bruijn indices) $\alpha_1 < \ldots < \alpha_m$.
We can construct an initial ucwf with $\vee$ and $(-)^+$, where the elements of $\ltm(n)$ are normal forms and $l = l' : \ltm(n)$ iff $l$ and $l'$ are identical normal forms.  If we extend the gat with a new sort $\leq$ and a new operator symbol $\refl$ for reflexivity, we can extend the initial ucwf with sets $\leq_n(l,l')$ that contain a single element $\refl(l)$ if $l = l' : \ltm(n)$ and are otherwise empty.

\paragraph{Remark on identity types.} The sort symbol $\leq$ for level equality resembles the identity type former $\Id$ in Martin-Löf type theory. We think of $\leq_n(l,l')$ as propositional level equality and its elements $p : \leq_n(l,l')$ as proofs of propositional level equality. However, while proofs $p : \Id_A(a,a')$ can make use of advanced logical reasoning, level equality proofs are limited and, as shown above, can only be obtained by equational reasoning from the laws for $\vee$ and $(-)^+$.

\ERASER{TC: Remark on finitely presented semilattices and constraints?}

\subsection{The level-indexed cwf of small types}\label{lcwf-sort}

We now define the gat of ucwf-indexed cwfs with extra structure for the small type formers $\Pi,\Sigma,\N_0,\N_1,\newline \N_2,\N,\W,\Id$.

Let $\L$ be the category of contexts in the ucwf of levels. We need to add sort symbols, operator symbols, and equations for the theory of presheaves
$$
T : \L^\mathrm{op} \to \CwF^{\Pi,\Sigma,\N_0,\N_1,\N_2,\N,\W,\Id}
$$
valued in the category of cwfs with extra structure for the small type formers and cwf-morphisms preserving cwf-structure and the structure of the small type formers strictly. Thus
\begin{itemize}
\item
$T(n)$ is the cwf (with extra structure) of contexts, substitutions, types, and terms that depend on level variables in $n$.
\item
Let $\sigma : n \to m$ be a level substitution. In the initial model it is an $m$-tuple of level terms in $n$ level variables, and
$T(\sigma) : T(m) \to T(n)$ substitutes the $m$ level variables by the respective $m$ level expressions in $\sigma$ in the various components of the cwf $T(m)$ yielding a cwf depending on $n$ level variables. All structure of the cwf with small type formers is preserved.
\end{itemize}

\paragraph{Sort symbols.} The gat for the level-indexed cwf of small types has the following sort symbols in addition to those of the gat of levels:
\begin{eqnarray*}
n : \lctx &\vdash& \ctx_n\\
n : \lctx, \Delta, \Gamma : \ctx_n &\vdash& \sub_n(\Delta,\Gamma)\\
n : \lctx, \Gamma : \ctx_n &\vdash& \ty_n(\Gamma)\\
n : \lctx, \Gamma : \ctx_n, A:\ty_n(\Gamma) &\vdash& \tm_n(\Gamma,A)
\end{eqnarray*}
These are the same as the sort symbols of the gat of cwfs, except that they are all indexed by an argument $n : \lctx$.

\paragraph{Operator symbols.} Similarly, the operator symbols are the same as for cwfs (with extra structure for the small type formers), except that they are also indexed by $n : \lctx$. The equations are modified accordingly. See the appendix.

The arrow part of the level-indexed cwf of small types axiomatizes level substitution. There is one operator symbol for each component of the level-indexed cwf. We overload notation:
\begin{eqnarray*}
n, n' : \lctx , \sigma : \lhom(n,n'), \Gamma : \ctx_{n'} &\vdash&
\Gamma[\sigma] : \ctx_n\\
n, n' : \lctx , \sigma : \lhom(n,n'), \Delta,\Gamma : \ctx_{n'}, \gamma : \sub_{n'}(\Delta,\Gamma)
&\vdash&
\gamma[\sigma] : \sub_{n}(\Delta[\sigma],\Gamma[\sigma]) \\
n, n' : \lctx , \sigma : \lhom(n,n'), \Gamma : \ctx_{n'}, A:\ty_{n'}(\Gamma)
&\vdash&
A[\sigma]: \ty_n(\Gamma[\sigma])\\
n,n' : \lctx , \sigma : \lhom(n,n'), \Gamma : \ctx_{n'}, A:\ty_{n'}(\Gamma), a : \tm_{n'}(A,\Gamma)
&\vdash&
a[\sigma] : \tm_{n}(A[\sigma],\Gamma[\sigma])
\end{eqnarray*}
\paragraph{Equations.}
The functor laws give us the following equations:
\begin{eqnarray*}
\Gamma[\lid_n] &=& \Gamma\\
\Gamma[\sigma \circ \tau] &=& \Gamma[\sigma][\tau]\\
\gamma[\lid_n] &=& \gamma\\
\gamma[\sigma \circ \tau] &=& \gamma[\sigma][\tau]\\
A[\lid_n] &=& A\\
A[\sigma \circ \tau] &=& A[\sigma][\tau]%: \ty_p(\Gamma[\sigma \circ \tau])
\\
a[\lid_n] &=& a\\
a[\sigma \circ \tau] &=& a[\sigma][\tau]%: \tm_p(A[\sigma \circ \tau],\Gamma[\sigma \circ \tau])
\end{eqnarray*}
Level substitution commutes with small type formers. We show the case for $\Pi$-types:

Let $n, n' : \lctx , \sigma : \lhom(n,n'), \Gamma : \ctx_{n'}, A:\ty_{n'}(\Gamma), B:\ty_{n'}(\Gamma.A)$. Then
\begin{eqnarray*}
\Pi(A,B)[\sigma] &=& \Pi(A[\sigma],B[\sigma])
\end{eqnarray*}
Moreover, if $b : \tm_n(\Gamma.A,B), c :  \tm_n(\Gamma,\Pi(A,B))$,  and $a : \tm_n(\Gamma,A)$, we have
\begin{eqnarray*}
\lambda(b)[\sigma] &=&\lambda(b[\sigma])\\
\app(c,a)[\sigma] &=&\app(c[\sigma],a[\sigma])
\end{eqnarray*}

\subsection{Level-indexed universes}

\paragraph{Operator symbols and a new sort symbol for level equality.}
We finally add the operator symbols and equations for level-indexed universes.
Each $T(n)$ has extra structure for level-indexed universes $\UU_l$ with decodings $\Ta_l$, where $l$ is a level term that depends on level variables in $n$. These universes are closed under the small type formers and contain smaller universes $\UU_{l'}$ for $l' < l$.

The operator symbols are obtained by internalizing the corresponding rules for the externally indexed universes.
\begin{eqnarray*}
l : \ltm(n), \Gamma : \ctx_n &\vdash& (\U_{l})_\Gamma : \ty_n(\Gamma)\\
l : \ltm(n), \Gamma : \ctx_n, a : \tm_n(\Gamma,(\U_{l})_\Gamma) &\vdash& {\Ta_{l}}(a) : \ty_n(\Gamma)\\
%\Gamma : \ctx &\vdash& (\N^0)_\Gamma : \tm(\Gamma,(\U_0)_\Gamma) \\
 l, l' : \ltm(n), \Gamma : \ctx_n,
a : \tm_n(\Gamma,(\U_{l})_\Gamma),
b :  \tm_n(\Gamma \cdot \Ta_{l}(a), (\U_{l'})_\Gamma)
&\vdash&
 \Pihat^{l,l'}(a,b) : \tm_n(\Gamma,(\U_{l \vee l'})_\Gamma)
\end{eqnarray*}
where we, as before, we have only showed closure under $\Pi$. Moreover, we have omitted the common premise $n : \lctx$ in each of the typings above.

As mentioned in Section \ref{ucwf-levels} we use level equality sorts when typing the operator symbols $\UU^m_l$ for universes $\UU_l$ in larger universes $\UU_m$:
% The following attempt
%\begin{eqnarray*}
%n : \lctx, l, m : \ltm(n), l < m, \Gamma : \ctx_n&\vdash&(\UU^m_l)_\Gamma: \tm_n(\Gamma,(\UU_{m})_\Gamma)
%\end{eqnarray*}
%where $l < m$ is defined as $l^+ \vee m = m$, does not work since equations are not allowed in argument types in gats.
%However, this issue can be resolved by instead using equality sorts as introduced above in Section
%\ref{ucwf-levels} and redefine $l < m$ as $\leq_n(l^+ \vee m, m)$. This yields the following proper typing of the operator symbol $\UU^m_l$:
\begin{eqnarray*}
n : \lctx, l, m : \ltm(n), p : l < m, \Gamma : \ctx_n&\vdash&(\UU^m_l)_{p,\Gamma} : \tm_n(\Gamma,(\UU_{m})_\Gamma)
\end{eqnarray*}
where $l < m$ is defined as $\leq_n(l^+ \vee m, m)$. In the sequel we will suppress the proof $p : l < m$ as an argument to this operator symbol and just write $(\UU^m_l)_{\Gamma}$.

\paragraph{Equations.}
The decoding equations for $\Ta_l$ and the equations for commutativity of operator symbols with substitution can be obtained by a straightforward internalization of the corresponding equations for the external tower. This means that the decoding equations are now relative to internal level contexts and level terms, as well as to terms. For example the decoding equation for $\Pi$
\begin{eqnarray*}
\Ta_{l \vee l'}(\Pi^{l,l'}(a,b)) = \Pi(\Ta_l(a),\Ta_{l'}(b)) :\tm_n(\Gamma,\UU_{l \vee l'})
\end{eqnarray*}
is now relative to the context
$$
n : \lctx, l, l' : \ltm(n), \Gamma : \ctx_n, a : \tm_n(\Gamma,\UU_l), b : \tm_n(\Gamma.\Ta_l(a),\UU_{l'} )
$$
and the decoding equation for the $l$th universe in the $m$th
\begin{eqnarray*}
\Ta_{m}((\UU^m_l)_\Gamma) &=& (\UU_l)_\Gamma
\end{eqnarray*}
is now relative to the context
$$
n : \lctx, l, m : \ltm(n), p : l < m
$$
where again $l < m$ is defined as $\leq_n(l^+ \vee m, m)$.

Equations for commutativity of operator symbols wrt term substitution $\gamma : \sub(\Delta,\Gamma)$:
%\begin{eqnarray*}
%\Ta^m_l(a) [ \gamma ] &=& \Ta^m_l(a[ \gamma ] )\\
%%\N^0_\Gamma [ \gamma ] &=&\N^0_\Delta\\
%\Pi^{l}(a,b)[ \gamma ] &=& \Pi^{l}(a [ \gamma ], b[ \gamma^\dagger ])
%\end{eqnarray*}
% where $\gamma : \sub(\Delta,\Gamma)$.
 \begin{eqnarray*}
(\U_l)_\Gamma [ \gamma ] &=& (\U_l)_\Delta\\
\Ta_l(a) [ \gamma ] &=& \Ta_l(a[ \gamma ] )\\
%\N^0_\Gamma [ \gamma ] &=&\N^0_\Delta\\
\Pi^{l,l'}(a,b)[ \gamma ] &=& \Pi^{l,l'}(a [ \gamma ], b[ \gamma^\dagger ])\\
(\UU^m_l)_\Gamma[ \gamma ] &=&(\UU^m_l)_\Delta
\end{eqnarray*}

Equations for commutativity of operator symbols wrt level substitution $\sigma: \lhom(n,n')$:  \begin{eqnarray*}
 (\U_{l})_\Gamma[\sigma] &=& (\U_{l[\sigma]})_{\Gamma[\sigma]}\\
 \Ta_l(a) [ \sigma ] &=& \Ta_{l[\sigma]}(a[ \sigma ] )\\
\Pi^{l,l'}(a,b)[ \sigma ] &=& \Pi^{l[ \sigma ] ,l'[ \sigma ] }(a [ \sigma ], b[ \sigma])\\
(\UU^m_l)_\Gamma[ \sigma ] &=&(\UU^{m[\sigma]}_{l[ \sigma ]} )_{\Gamma[ \sigma ]}
 \end{eqnarray*}

\subsection{Cumulativity}\label{cumulativity}
\paragraph{Operator symbol.}
An operator symbol for cumulativity is obtained by internalizing the operator symbols for cumulativity in the external tower:
\begin{eqnarray*}
n : \lctx, l,m : \ltm(n), p : l < m, \Gamma : \ctx_n, a : \tm_n(\Gamma,(\U_{l})_\Gamma) &\vdash& {\Ta^m_{l}}(a) : \tm_n(\Gamma,(\U_{m})_\Gamma))
\end{eqnarray*}
where again $l < m$ is defined as $\leq_n(l^+ \vee m, m)$. We have the equations:
\begin{eqnarray*}
\Ta_m(\Ta^m_{l}(a)) &=& \Ta_l(a)\\
\Ta^m_{l}(a)[\gamma] &=& \Ta^m_{l}(a[\gamma])
\end{eqnarray*}
In the presence of cumulativity we can replace the doubly indexed codes for $\Pi$ by the following singly indexed version:
\begin{eqnarray*}
\Gamma : \ctx,
a : \tm(\Gamma,(\U_{l})_\Gamma),
b :  \tm(\Gamma \cdot \Ta_{l}(a), (\U_{l})_\Gamma))
&\vdash&
 \Pihat^{l}(a,b) : \tm(\Gamma,(\U_{l})_\Gamma)
 \end{eqnarray*}
\paragraph{Equations.}
We have the following decoding equation:
\begin{eqnarray*}
\Ta^{m}_{l}(\Pi^{l}(a,b)) = \Pi^{m}(\Ta^m_l(a),\Ta^{m}_{l}(b)) &:& \tm_n(\Gamma,(\UU_{m})_\Gamma)
\end{eqnarray*}
where $n: \lctx,l,m : \ltm(n), p: l < m, \Gamma : \ctx_n, a : \tm_n(\Gamma,(\UU_l)_\Gamma), b : \tm_n(\Gamma.\Ta_l(a),(\UU_{l})_\Gamma)$.
The decoding equation for lifting of codes for universes is as follows:
\begin{eqnarray*}
n : \lctx, k, l, m : \ltm(n), p : k < l, q : l < m, \Gamma : \ctx(n) &\vdash& \Ta^m_l((\UU^l_k)_\Gamma) = (\UU^m_k)_\Gamma
\end{eqnarray*}
Equations for commutativity of lifting and term substitution $\gamma : \sub(\Delta,\Gamma)$:
\begin{eqnarray*}
\Ta^m_l(a) [ \gamma ] &=& \Ta^m_l(a[ \gamma ] )
%\N^0_\Gamma [ \gamma ] &=&\N^0_\Delta\\
%\Pi^{l}(a,b)[ \gamma ] &=& \Pi^{l}(a [ \gamma ], b[ \gamma^\dagger ])
\end{eqnarray*}
and level substitution $\sigma : \lhom(n,n')$:
\begin{eqnarray*}
\Ta^m_l(a) [ \sigma ] &=& \Ta^m_{l[\sigma]}(a[ \sigma ] )
\end{eqnarray*}
%Check that this is right. Cf comment by intern.

\subsection{Level-indexed products of types}
%We add three operator symbols in addition to the operator symbols for cwfs in Section 5.2 and 5.3:
\def\l{\mathrm{l}}
%\begin{tiny}
In \cite[Section 4]{BezemCDE22} we introduced
universal level quantification $[\alpha]A$ with level abstraction $\tuple{\alpha}a$ and application $a\,l$ of a term to a level. The respective operator symbols are $\forall_\l, \lambda_\l,$ and $\app_\l$:
\begin{eqnarray*}
n : \lctx, \Gamma : \ctx_n, B : \ty_{\s(n)}(\Gamma[\lp])&\vdash& \forall_\l(B) : \ty_n(\Gamma)\\
n : \lctx, \Gamma : \ctx_n, B : \ty_{\s(n)}(\Gamma[\lp]), b : \tm_{\s(n)}(\Gamma[\lp], B) &\vdash& \lambda_\l(b) : \tm_n(\Gamma,\forall_\l(B))\\
n : \lctx, \Gamma : \ctx_n, B : \ty_{\s(n)}(\Gamma[\lp]), c :  \tm_n(\Gamma,\forall_\l(B)), l : \ltm(n) &\vdash& \app_\l(c,l) : \tm_n(\Gamma, B[\tuple{\lid,l}])
\end{eqnarray*}
Equations ($\beta$ and $\eta$ for level abstraction and application):
 \begin{eqnarray*}
 \app_\l(\lambda_\l(b),l) &=& b[\tuple{\lid,l}]\\
 \lambda_\l(\app_\l(c[\lp],\lq)) &=& c
 \end{eqnarray*}
 Equations for commutativity of operator symbols with respect to term and level substitution:
 \begin{eqnarray*}
 \forall_\l(B)[ \gamma ] &=& \forall_\l(B[ \gamma[\lp]])\\
 \lambda_\l(B)[ \gamma ] &=& \lambda_\l(B[ \gamma[\lp]])\\
 \app_\l(c,l)[ \gamma ] &=& \app_\l(c[ \gamma ] ,l[ \gamma ] )\\
 \forall_\l(B)[ \sigma ] &=& \forall_\l(B[ \sigma^{\dagger}])\\
 \lambda_\l(B)[ \sigma ] &=& \lambda_\l(B[ \sigma^{\dagger}])\\
\app_\l(c,l)[ \sigma ] &=& \app_\l(c[ \sigma ] ,l[ \sigma ] )
\end{eqnarray*}
where $\gamma : \sub_n(\Delta,\Gamma)$ is a term substitution, and $\sigma : \lhom(m,n)$ is a level substitutuion with $\sigma^\dagger = \tuple{\sigma \circ \lp, \lq}$.
To check the type of the equation
$$ \forall_\l(B)[ \sigma ] = \forall_\l(B[ \sigma^{\dagger}])$$ we assume $\Gamma : \ctx_n, B : \ty_{\s(n)}(\Gamma[\lp])$. It follows that $B[\sigma^\dagger] : \ty_{s(m)}(\Gamma[\lp][\sigma^\dagger]) = \ty_{s(m)}(\Gamma[\sigma][\lp])$. Hence
$\forall_\l(B[ \sigma^{\dagger}]) : \ty_m(\Gamma[\sigma])$.

We refer the reader to the appendix for the remaining rules of $\Sigmaint$.

\paragraph{Remark on universal quantification in predicate logic.}

In the introduction we mentioned that untyped predicate logic can be captured by ucwf-indexed scwfs. (This gives a proof-relevant notion of model, like Lawvere's hyperdoctrines \cite{lawvere:hyperdoctrines}.) We note that the operator symbols and equations for universal quantification in untyped predicate logic are the same as those for universal level quantification above, except the difference between scwfs and cwfs: propositions in predicate logic do not depend on proofs.

\subsection{Syntax and inference rules as an initial model}\label{initiality-internal}

\paragraph{A generic construction based on explicit substitution.} The gat $\Sigmaint$ for level-indexed cwfs, explicit universe polymorphism, and level-indexed products is finitary. Hence we can directly instantiate the construction of the term model in \cite{bezem:hofmann} and get a proof that $\T_{\Sigmaint}$ is initial in $\CwF_{\Sigmaint}$, the category of cwfs with an {\em internal} level-indexed cwf with extra structure. In a similar way as we showed in Section \ref{initiality-external} an object in $\CwF_{\Sigmaint}$ can be externalized yielding an object in the category $\LCwFint$ of level-indexed cwfs with extra structure for explicit universe polymorphism, and level-indexed products.

\paragraph{A construction based on implicit substitution (initiality conjecture).} We shall outline a construction of an initial object $\Tint = (\Lctx,\Lsub,\Ltm,\Ctx,\Sub,\Ty,\Tm)$ in $\CwFint$ based on $\TTint$, the decorated version of our type theory with explicit universe polymorphism \cite[Section 4]{BezemCDE22}.
%In this theory level variables and ordinary term variables can appear in any order. As already mentioned we consider, without loss of generality, a version of $\TTint$ where level variables precede ordinary term variables (both using de Bruijn indices). In this theory every context has the form $n,\Gamma$, where $n$ is the number of available level variables, and $\Gamma$ is a context of term variables that may depend on those $n$ level variables.
We need to show (i) how to organize the level terms and level judgments into a ucwf of levels; (ii) how to construct a level-indexed cwf of small types by modifying the construction of a cwf of small types; and (iii) how to construct a model of the level-indexed universes by modifying the construction of a model of the externally indexed universes based on $\TText$.

We first construct a ucwf $(\Lctx,\Lsub,\Ltm)$ of levels. To this end we extend Brilakis'  \cite{Brilakis18} construction in Agda of the equivalence between two initial ucwfs: one with explicit substitution and one with implicit substitution (defined by recursion on terms) and de Bruijn variables. Since the ucwf with explicit substitution is constructed directly from the operator symbols of ucwfs, Brilakis' proof is essentially the same as proving the initiality of the ucwf with implicit substutions. To prove the initiality of the ucwf of levels, we extend Brilakis' proof with the extra structure for $\vee$ and $(-)^+$.
\begin{itemize}
\item An element $n \in \Lctx$ is the number of available level variables.
\item An element of $\Ltm(n)$ is an equivalence class of level terms generated by $\vee$ and $(-)^+$ from $n$ level variables with respect to the equivalence relation generated by the equations for $\vee$ and $(-)^+$. In our type theory with explicit universe polymorphism \cite{BezemCDE22} this corresponds to the level terms $l$ such that
$$
n, \Gamma \vdash l\ \level
$$
and two terms $l, l' : \ltm(n)$ are equivalent provided
$$
n, \Gamma \vdash l = l'
$$
\item An element of $\Lsub(m,n)$ is an equivalence class of sequences of level terms.
\item We refer to Brilakis for the definition of the ucwf-operations. Note that level substitution $l[\sigma]$ is defined by induction on $l$. The definition of $\vee$ and $(-)^+$ on equivalence classes of levels is immediate.
% If $n = (\alpha_n, \ldots, \alpha_1)$ and $\s(n) =  (\alpha_n, \ldots, \alpha_1, \alpha_0)$ are lists of distinct level variables, then $\lp_n = (\alpha_n, \ldots, \alpha_1) : \lhom(\s(n),n)$ and $\lq_n = \alpha_0 : \ltm(\s(n))$ are the projections.
\end{itemize}

The next step is to construct the level-indexed cwfs (with extra structure) $(\Ctx_n,\Sub_n,\Ty_n,\Tm_n)$ for $n \in \Lctx$:
\begin{itemize}
\item $\Ty_n(\Gamma)$ is the set of equivalence classes of raw types $A$ such that $n, \Gamma \vdash A$ is a type and $A.A'$ are equivalent provided $n, \Gamma \vdash A = A'$.
\item $\Tm_n(\Gamma,A)$ is the set of equivalence classes of raw terms $a$ such that $n, \Gamma \vdash a : A$ and $a.a'$ are equivalent provided $n, \Gamma \vdash A = A'$.
\item $\Ctx_n$ is the set of equivalence classes of raw contexts $\Gamma$ such that $n, \Gamma \vdash$ under the equivalence relation $n, \Gamma = \Gamma' \vdash$.
\item $\Sub_n(\Delta,\Gamma)$ is the set of equivalence classes of raw context morphisms.
\end{itemize}
We then define all the operator symbols in this structure and check the equations.

The final part of the construction is to interpret the operator symbols for level substitution and composition (in levels, level morphisms, contexts, context morphisms, types and terms) that correspond to the arrow part of the level-indexed cwf with extra structure. These are all defined implicitly by induction on the raw syntax.

This concludes the outline of the construction of an object of $\CwFint$. Finally, we need to construct a morphism to any other object in $\CwFint$ and prove that this is unique.

\section{Conclusion}

We presented an infinitary gat $\Sigmaext$ for Martin-Löf type theory with an external tower of universes and a finitary gat $\Sigmaint$ for Martin-Löf type theory with internally level-indexed universes and level-indexed products. We have also explained that the models $\T_{\Sigmaext}$ and $\T_{\Sigmaint}$ are instances of general constructions of initial models of gats in our article
\cite{bezem:hofmann}. Moreover, we outlined the constructions of the {\em external} cwfs $\Text$ and $\Tint$, initial in $\CwFext$ and $\CwFint$ respectively.

A key ingredient of the gat $\Sigmaint$ is the sort symbol for level equality $\leq$. In a forthcoming article we plan to show how level equality sorts can be employed for representing equational constraints. This will enable us to extend $\Sigmaint$ with new sort symbols and equations for the extension of $\TTint$ with equational constraints presented in \cite[Section 5]{BezemCDE22}.

%We would also like to spell out the externalization process mentioned in Section \ref{initiality-external} mapping for example a cwf with an internal cwf to the corresponding cwf, and uniformly for other gats.

%We note that in cwfs (and scwfs, ucwfs, and indexed cwfs) substitution is modelled by the arrow part of a functor. Under certain conditions we can define the corresponding substitution combinators in the term model by induction. This relies on identifying variables (projections) as a special subclass of the terms in a cwf (and the like). It seems that in this way we get a uniform way of constructing term models for various type theories and predicate logics which are close to the standard presentations. This is also a topic for future research.
%
\paragraph{Related research.}
An alternative approach to representing type theories is in terms of a {\em Logical Framework}, such as Martin-Löf's
\cite{nordstrom:book}, Edinburgh LF \cite{harper-honsell-plotkin} or Dedukti \cite{dowek-dedukti}. These are based on dependent type theories with $\Pi$-types and one or more universes. The aim is to encode other logics by adding constants and equations to the logical framework. We contrast this to gats which are based on dependent types {\em without} $\Pi$-types and universes. Logics are then encoded by adding sort symbols, operator symbols, and equations to the basic theory of dependent types.

Logical frameworks have received renewed interest as an approach to Voevodsky's initiality conjecture project. Examples include the work by Bauer, Haselwarter, and Lumsdaine \cite{BauerHL20}, Uemura \cite{Uemura23}, and Kaposi and Xie \cite{kaposi:sogat} on second-order generalized algebraic theories (SOGATs). These references present several examples of encodings of theories, including basic dependent type theory, 2-level type theory, predicate logic, and cubical type theories. It seems likely that our type theory with universe polymorphism could be encoded compactly in a similar way. Kaposi and Xie also propose a general translation from their SOGATs to GATs. However, we leave the SOGAT-encoding of our theories to future work and also the question of the relationship between the gat in our paper and the one obtained by applying the translation from SOGATs to GAts.

Another line of related research is on the quotient inductive-inductive types (qiits) of Kaposi, Kov{\'{a}}cs, and Altenkirch \cite{kaposi:qiits,kovacs:phd} that are closely related to gats. Although formal details differ, qiits are roughly initial gats considered as data types in dependent type theory. Sort symbols correspond to data type constructors; operator symbols correspond to term constructors; and equations between terms can be declared. Since a qiit is inductively generated, it has an elimination principle. It is the latest in the following sequence of more and more general inductive notions in dependent type theory: inductive type, inductive family, inductive-inductive type, and quotient inductive-inductive type.

%and $\gamma^\dagger = \tuple{\gamma \circ \p, \q}$??
%\end{tiny}
%\footnote{Discussion with Raphael Sterback:
%$$
%\forall_l (B)[\gamma] = \forall_l (B[\gamma[\lp]])
%$$}

\bibliographystyle{eptcs}
\bibliography{localrefs}
\appendix
\section{Operator symbols and equations for level-indexed cwfs of small types}

We already listed the sort symbols in \ref{lcwf-sort}.

\paragraph{Operator symbols.}
%\footnote{Check whether implicit arguments are systematic.}
%We have the following sort symbols:
%\begin{eqnarray*}
%n : \lctx &\vdash& \ctx_n\\
%n : \lctx, \Delta, \Gamma : \ctx_n &\vdash& \sub_n(\Delta,\Gamma)\\
%\end{eqnarray*}
\begin{eqnarray*}
n : \lctx, \Gamma : \ctx_n &\vdash& \id_{n,\Gamma} : \sub_n(\Gamma,\Gamma)\\
n : \lctx, \Xi,\Delta,\Gamma : \ctx_n, \gamma : \sub_n(\Delta,\Gamma), \delta : \sub_n(\Xi,\Delta) &\vdash&
\gamma \circ \delta : \sub_n(\Xi,\Gamma)\\
n : \lctx, \Gamma,\Delta : \ctx_n, A:\ty_n(\Gamma), \gamma : \sub_n(\Delta,\Gamma) &\vdash&
A[\gamma] : \ty_n(\Delta)\\
n : \lctx, \Gamma,\Delta : \ctx_n, A:\ty_n(\Gamma), \gamma : \sub_n(\Delta,\Gamma), a:\tm_n(\Gamma,A) &\vdash&  a[\gamma] : \tm_n(\Delta,A[\gamma])\\
n : \lctx &\vdash& 1_n : \ctx_n\\
n : \lctx, \Gamma : \ctx_n &\vdash& \tuple{}_{n,\Gamma} : \sub_n(\Gamma,1_n)\\
n : \lctx, \Gamma : \ctx_n, A:\ty_n (\Gamma) &\vdash& \Gamma \cext A : \ctx_n \\
n : \lctx, \Gamma,\Delta : \ctx_n , A:\ty_n (\Gamma), \gamma : \sub_n (\Delta,\Gamma), a:\tm_n (\Delta,A[\gamma]) &\vdash& \tuple{\gamma,a} : \sub_n (\Delta,\Gamma\cext A)\\
n : \lctx, \Gamma : \ctx_n , A:\ty_n (\Gamma) &\vdash& \p_{n,\Gamma,A}: \sub_n (\Gamma\cext A,\Gamma)\\
n : \lctx, \Gamma : \ctx_n , A:\ty_n (\Gamma) &\vdash& \q_{n,\Gamma,A}: \tm_n (\Gamma\cext A,A[\p])
\end{eqnarray*}
Operator symbols for level-indexed $\Pi$-types
%(should we decorate $\lambda$ and $\app$ with type information?):
\begin{eqnarray*}
n : \lctx, \Gamma : \ctx_n, A : \ty_n(\Gamma), B : \ty_n(\Gamma.A)&\vdash& \Pi(A,B) : \ty_n(\Gamma)\\
n : \lctx, \Gamma : \ctx_n, A : \ty_n(\Gamma), B : \ty_n(\Gamma.A), b : \tm_n(\Gamma.A, B) &\vdash& \lambda(b) : \tm_n(\Gamma,\Pi(A,B))\\
n : \lctx, \Gamma : \ctx_n, A : \ty_n(\Gamma), B : \ty_n(\Gamma.A), c :  \tm_n(\Gamma,\Pi(A,B)), a : \tm_n(\Gamma, A) &\vdash& \app(c,a) : \tm_n(\Gamma, B[\tuple{\id,a}])
\end{eqnarray*}

\paragraph{Equations.}
\begin{eqnarray*}
\id_{n,\Gamma} \circ \gamma &=& \gamma \\
 \gamma \circ \id_{n,\Delta} &=& \gamma \\
(\gamma \circ \delta) \circ \xi &=& \gamma \circ (\delta \circ \xi)\\
%&\vdash&
A[\id_{n,\Gamma}] &=& A
%: \ty_n(\Gamma)
\\
%a:\tm_n(\Gamma,A) &\vdash&
a[\id_{n,\Gamma}] &=& a% : \tm_n(\Gamma,A)
\\
%n : \lctx, \Xi,\Delta,\Gamma : \ctx_n, \delta : \sub_n(\Xi,\Delta), \gamma : \sub_n(\Delta,\Gamma),
%&\vdash&
A[\gamma\circ\delta] &=& A[\gamma][\delta]
%: \ty_n(\Xi)
\\
%n : \lctx, \Xi,\Delta, : \ctx_n, \delta : \sub_n(\Xi,\Delta), \gamma : \sub_n(\Delta,\Gamma),
%a:\tm_n(\Gamma,A) &\vdash&
a[\gamma\circ\delta] &=& a[\gamma][\delta]
%: \tm_n(\Xi,A[\gamma\circ\delta])
\\
%n : \lctx &\vdash&
\id_{n,1_n} &=& \tuple{}_{n,1_n}
%: \sub_n(1_n,1_n)
\\
%\Gamma,\Delta : \ctx_n , \gamma : \sub_n (\Delta,\Gamma) &\vdash&
\tuple{}_{n,\Gamma}\circ\gamma &=& \tuple{}_{n,\Delta}% : \sub_n (\Delta,1_n )
\\
%\Gamma,\Delta : \ctx, A:\ty(\Gamma), \gamma : \sub(\Delta,\Gamma), a:\tm(\Delta,A[\gamma]) &\vdash&
\p_{n,\Gamma,A}\circ\tuple{\gamma,a} &=& \gamma : \sub(\Delta,\Gamma)\\
%\Gamma,\Delta : \ctx, A:\ty(\Gamma), \gamma : \sub(\Delta,\Gamma), a:\tm(\Delta,A[\gamma]) &\vdash&
\q_{n,\Gamma,A}[\tuple{\gamma,a}] &=& a : \tm(\Delta,A[\gamma]) \\
%\Gamma,\Delta,\Xi : \ctx, A:\ty(\Gamma), \gamma : \sub(\Delta,\Gamma), a:\tm(\Delta,A[\gamma]), \delta : \sub(\Xi,\Delta) &\vdash&
\tuple{\gamma,a} \circ \delta &=& \tuple{\gamma\circ\delta,a[\delta]} \\
%:
%\sub(\Xi,\Gamma\cext A) \\
%\Gamma : \ctx, A:\ty(\Gamma) &\vdash&
\id_{n,\Gamma\cext A} &=& \tuple{\p_{n,\Gamma,A},\q_{n,\Gamma,A}} : \sub(\Gamma\cext A,\Gamma\cext A)
\end{eqnarray*}
\ERASER{MB: check. Perhaps better to describe how to modify \ref{sec:gatPiU}:
prefix every rule with $n:\lctx$ and add index $n$ to every occurrence
of $\ctx,\ty,\tm$. PD: we didn't spell all of this out in  \ref{sec:gatPiU}. This is here, since we want to have a complete definition of $\Sigmaint$.}
Equations (omitting the context and type of the equalities):
 \begin{eqnarray*}
 \app(\lambda(b),a) &=& b[\tuple{\id,a}]\\
 \lambda(\app(c[\p],\q)) &=& c
 \end{eqnarray*}
 Equations for commutativity of operator symbols wrt substitution:
 \begin{eqnarray*}
\Pi(A,B)[ \gamma ] &=& \Pi(A [ \gamma ], B[ \gamma^\dagger ])\\
\lambda(b) [ \gamma ] &=& \lambda(b[\gamma^\dagger ])\\
\app(c,a) [ \gamma ] &=& \app(c[ \gamma ], a[ \gamma ] )
\end{eqnarray*}
where $\gamma^\dagger = \tuple{\gamma \circ \p, \q}$.

\end{document}